\documentclass[aps, prl, twocolumn, superscriptaddress, showpacs, preprintnumbers]{revtex4-1}

\usepackage{amssymb, amsmath, bm, float, epsfig, slashed}
\usepackage{color}
\usepackage[centering]{geometry}
\begin{document}
\preprint{}
\title{(De)Constructing magnetized dimensions}
\author{Yoshiyuki~Tatsuta}
\email{y\_tatsuta@akane.waseda.jp}
\affiliation{Department of Physics, Waseda University, Tokyo 169-8555, Japan}
\author{Akio~Tomiya}
\email{akio.tomiya@mail.ccnu.edu.cn}
\affiliation{Key Laboratory of Quark \& Lepton Physics (MOE) and Institute of Particle Physics,\\ Central China Normal University, Wuhan 430079, China}

\date{\today}

\begin{abstract}
We provide an origin of family replications in the standard model of particle physics by constructing renormalizable, asymptotically free, four dimensional local gauge theories that dynamically generate the fifth and sixth dimensions with magnetic fluxes. 
\end{abstract}

\maketitle

{\em Introduction.}--- 
The standard model (SM) of elementary particle physics still holds a mysterious puzzle in its matter contents.
{\em Who ordered three copies of quarks and leptons in the SM?}
As I.\,I.\,Rabi famously quipped for the muon {\em ``Who ordered that?''}, up-type quarks (up $u$, charm $c$, top $t$), down-type quarks (down $d$, strange $s$, bottom $b$), charged leptons (electron $e$, muon $\mu$, tauon $\tau$) and neutrinos (electron neutrino $\nu_e$, muon neutrino $\nu_\mu$, tau neutrino $\nu_\tau$) carry the same quantum charges and are distinguished by their Yukawa couplings to the Higgs boson, namely their masses.
After the SM was proposed, this simple and profound mystery has been one of great interests in the SM over a long period of history.
Nevertheless, a satisfactory explanation for an origin of three-generation structures has not been naturally given from the viewpoint of four dimensional (4D) field theory.

Among attempts to reveal an origin of triply multiple copies of the SM fermions, a promising hypothesis is that there exist compactified extra spacial dimensions in addition to 4D that we live in, although our world apparently looks four dimensional.
In higher dimensional theories, fields are expanded by Kaluza--Klein (KK) expansions, which are to decompose extra dimensional parts of higher dimensional fields into a complete set spanned by the KK mode wavefunctions.
A novel proposal by Bachas is based on quantized magnetic fluxes in toroidal compactification \cite{Bachas:1995ik}, where the magnetic fluxes provide a degeneracy of the lowest KK-expanded wavefunctions. 
The degeneracy of the lowest KK modes should be regarded as the family replication of matters after dimensional reduction.
After the Bachas's proposal, Ibanez et al. pointed out that the degenerated mode functions can be analytically expressed by an elliptic function and discussed their convergence properties \cite{Cremades:2004wa}.

However, such higher dimensional theories contain dimensionful coupling constants and are non-renormalizable.
Thus, these theories possess less predictability in general.
In 2001, Arkani-Hamed et al. proposed a splendid ultraviolet (UV) completion, called {\em (de)construction} of an extra dimension \cite{ArkaniHamed:2001ca}, where a fifth dimension can be effectively established by the multiplicity of 4D renormalizable gauge theories.
In the paper, the authors utilized the knowledge of lattice gauge theories to interpret the non-renormalizable higher dimensional gauge theories by renormalizable ones.
As the results, (de)constructed higher dimensional theories acquire a predictability without loss of several essential properties.

In this Letter, we apply the (de)construction mechanism to the flux compactification where there exists a (nontrivial) topological index associated with the index theorem, and construct renormalizable, asymptotically free, 4D local gauge theories that dynamically generate the fifth and sixth dimensions with magnetic fluxes.
A main subject of this study is to establish the UV completion of magnetized toroidal compactifications and formulate latticized gauge theories where the index theorem \cite{Atiyah:1963zz} is applicable.
We find that two dimensional Moose diagram is insufficient due to the Nielsen--Ninomiya theorem \cite{Nielsen:1981hk} with the index theorem for zero-modes where the index of the Dirac operator in the latticized space is inevitably zero.
Thus, it is remarkable that it is necessary to treat three dimensional (3D) Moose diagram to correctly regulate the theory from (de)constructing points of view.

\vspace{5pt}
{\em Multiple zero-modes with fluxes.}--- 
Here, we briefly review an original theory proposed in Ref.\,\cite{Cremades:2004wa}.
We consider a six dimensional (6D) gauge theory that has magnetic fluxes (or magnetic monopoles) in toroidal compactification $T^2$.
It contains nonvanishing flux background $b = \int_{T^2} F$ of the field strength $F=(ib/2) dz \wedge d \bar z$, which is provided by a vector potential
\begin{gather}
A^{(b)}(z) = \frac{b}2 {\rm Im} \, (\bar z dz),
\label{bgd}
\end{gather}
where two Cartesian coordinates of the fifth and sixth directions $y_5$ and $y_6$ are expressed by $z \equiv (y_5 + i y_6)/2 \pi R$.

The single-valuedness of the 6D action under contractible loops, e.g., $z \to z+1 \to z+1+i \to z + i \to z$, demands the Dirac charge quantization,
\begin{gather}
\frac{qb}{2\pi} = m \in \mathbb{Z}.
\end{gather}
We perform the KK decomposition of the 6D Weyl spinor $\Psi$ and scalar $\Phi$ by a complete set as 
\begin{align}
\Psi(x^\mu, z) &=\sum_n \chi_n(x^\mu) \otimes \psi_n(z),\\
\Phi(x^\mu, z) &= \sum_n \varphi_n(x^\mu) \otimes \phi_n(z).
\end{align}
for $n=0,1,2, \cdots$, where $\chi_n$ denotes 4D Weyl spinors.
Here, the KK-decomposed wavefunctions in the extra dimensional parts $\psi_n$ and $\phi_n$ are chosen to be eigenstates of the covariant derivative $D = 2 \partial/\partial \bar z + \pi m z$ and the Laplace operator $\Delta = \{D^\dag, D\}/2$ for extra dimensions as $i \slashed{D} \psi_n = m_n \psi_n$ and $\Delta \phi_n = m_n^2 \phi_n$, respectively, as defined in Ref.\,\cite{Cremades:2004wa}.
In the Letter, we adopt a unit $2 \pi R =1$ for the compactification radius $R$ and set a $U(1)$ charge as $q=1$.
On the flux background \eqref{bgd} for $m > 0$, the lowest KK-decomposed modes $\psi_0$ are $m$-multiply degenerated as
\begin{gather}
\psi^j_{+, 0}(z) = {\cal N} e^{\pi i m z {\rm Im} \, z} \, \vartheta
\begin{bmatrix}
j/m\\[5pt]
0
\end{bmatrix}
(mz, mi), \label{eq:eigenvalue_CIM_eq}
\end{gather}
with $j=0,1, \cdots, m-1$ and the Jacobi theta function
\begin{gather}
\vartheta
\begin{bmatrix}
\alpha\\[5pt]
\beta
\end{bmatrix}
(\nu, \tau) = \sum_{\ell \in \mathbb{Z}} e^{\pi i (\alpha+\ell)^2 \tau + 2\pi i(\alpha + \ell)(\nu +\beta)},
\end{gather}
where $\alpha$ and $\beta$ are real parameters, and $\nu$ and $\tau$ take complex values with ${\rm Im} \, \tau>0$.
On another hand, $\psi_{-, 0}$ possesses no normalizable zero-mode wavefunction, where we decompose the two dimensional (2D) spinor as $\psi_0 = (\psi_{+, 0}, \psi_{-, 0})^T$ carrying the 2D chiralities. 
Also, ${\cal N}$ denotes a normalization factor, given as ${\cal N} = (2m)^{1/4}$.
Note that the lack of $\psi_{-, 0}$ implies that the chiral spectra are realizable in the low energy effective theory.
The zero-modes \eqref{eq:eigenvalue_CIM_eq} are localized at different regions on the torus $T^2$ and their schematic shapes are Gaussian-like.
The important point is that the degeneracy of the zero-modes corresponds to the family replication after dimensional reduction.
Although mode functions of the scalar are the same as those of the spinor, an exception remains in the KK mass spectra.
Indeed, it is straightforwardly found that the KK mass spectra of the spinor are calculated as 
\begin{gather}
m^2_n = 4 \pi m n. \label{eq:KK_mass}
\end{gather}
(The KK mass spectra of the scalar are given as $m^2_n = 4 \pi m (n+1/2)$.) 
Although they are not necessary in this Letter, the concrete wavefunctions of excited KK modes $\psi_n$ and $\phi_n$ for $n\geq 1$ are analytically calculated, as discussed in Ref.\,\cite{Hamada:2012wj}.

\vspace{5pt}
{\em (De)Construction of magnetized dimensions.}---
\begin{figure}[t]
\centering
\includegraphics*[width=5cm]{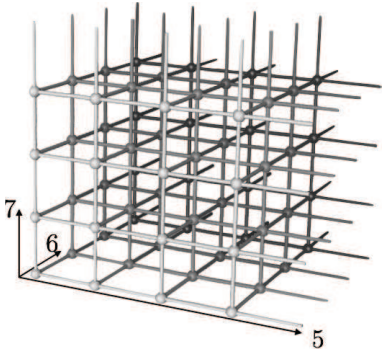}
\caption{A schematic picture of three dimensionally extended Moose diagram (or discretized toroidal extra directions plus another direction associated with the domain-wall fermion), which is necessary to utilize the index theorem. Link functions $Q_M(\bar{y})$ with $\bar{y}=(y_5,y_6,y_7)$ and $M=5,6,7$, correspond to gauge fields in the lattice gauge theory.}
\label{fig:6d_lat}
\end{figure}
In order to (de)construct magnetized toroidal dimensions, we start from the Wilson fermion on the 3D ``Moose'' diagram in Fig.\,\ref{fig:6d_lat} or namely discretized three dimensions.
The parts associated with the seventh dimension of fields satisfy the Dirichlet boundary condition.
The Lagrangian of the 4D Weyl spinor in three Cartesian coordinates reads:
\begin{widetext}
\begin{align}
{\cal L}
&= \frac{1}{2}\sum_{\bar{y}} \bar{\eta}_{\bar{y}} i \left[\sum_{M=5,6,7} \Gamma_M\big(Q_{M}(\bar{y}) \eta_{\bar{y}+\hat{M}} - Q^\dagger_{M}(\bar{y}-\hat{M} ) \eta_{\bar{y}-\hat{M}} \big) \right] - M_0 \sum_{\bar{y}} \bar{\eta}_{\bar{y}} \eta_{\bar{y}}\notag\\
& \quad\qquad - \frac{1}{2}\sum_{\bar{y}} \bar{\eta}_{\bar{y}} \left[\sum_{M=5,6,7} \big(Q_{M}(\bar{y}) \eta_{\bar{y}+\hat{M}} 
+ Q^\dagger_{M}(\bar{y}-\hat{M} ) \eta_{\bar{y}-\hat{M}} -2 \bar{\eta}_{\bar{y}} \eta_{\bar{y}}\big) \right],
\end{align}
\end{widetext}
where $\bar{y} \equiv (y_5, y_6, y_7)$ is a discretized coordinate for the 3D Moose diagram which is defined as $y_M \in \{1/N_M, 2/N_M, \cdots, 1\} \,\, (M=5,6,7)$, $\hat{M}$ is a unit vector for the $M$-th direction and $\Gamma_M$ is the gamma matrices for extra three dimensions.
As is the case with the lattice gauge theory, $\eta_{\bar{y}} = {\eta}_{\bar{y}}(x)$ and $\bar{\eta}_{\bar{y}} = \bar{\eta}_{\bar{y}}(x)$ are regarded as independent fermion fields each other in the 4D Minkowski spacetime, and $M_0$ is a real positive parameter of order one.
Since we are interested in the low energy spectrum of the Dirac operator, thus we extract the low lying modes of the Dirac operator as the same as the work in Ref.\,\cite{Tomiya:2016jwr}.
To be precise, after integrating out only the seventh direction of the latticized fermion fields with a trivial link function $Q_7(\bar{y})=1$ and a Pauli--Villars infrared regulator field, chiral fermions are realized in the residual 2D, similarly to a derivation of the overlap-Dirac operator in the context of the 4D lattice gauge theory.
For practical reasons, we employ the M\"obius domain-wall Dirac operator rather than the standard domain-wall Dirac operator.
This is due to the fact that the M\"obius domain-wall Dirac operator can realize more appropriate chiral symmetry \cite{Brower:2005qw} only through a slight extension of the seventh direction in comparison with the standard one, e.g., Ref.\,\cite{Fukaya:2004kp}.
Also, it is expected that the M\"obius domain-wall fermion with $N_7$ lattice points becomes asymptotically equivalent to the standard one with $2N_7$ lattice points \cite{Brower:2012vk}.
The Hermitian domain-wall Dirac operator $H_\text{DW}$ \cite{Kaplan:2009yg} with the corresponding Pauli--Villars infrared regulator filed is given in Refs.\,\cite{Brower:2005qw, Brower:2012vk} as
\begin{gather}
\label{eq:def_dw_eff}
H_\text{DW}=\frac{1}{2}\text{sgn}_\text{rat}(H_M),
\end{gather}
where the matrix sign function $\text{sgn}_\text{rat}(H_M)$ by use of the rational approximation is defined as 
\begin{gather}
\label{eq:polar}
\text{sgn}_\text{rat}(H_M)=\frac{1-(T(H_M))^{N_7}}{1+( T(H_M))^{N_7}},
\end{gather}
with the transfer matrix $T(H_M)=(1-H_M)/(1+H_M)$.
The kernel operator (the Hermitian M\"obius-Dirac operator) $H_M$ is given as
\begin{gather}
\label{eq:dw_kernel}
H_M=\Gamma_7\frac{2 D_W}{2+D_W},
\end{gather}
where $\Gamma_7$ is a chirality operator, namely the Pauli matrix $\sigma_3$, and $D_W$ is the Wilson--Dirac operator with the mass $-M_0$. 
Here, $0 < M_0 < 2$ is required to realize a correct pole structure, and we set $M_0 = 1$.
We adopt the same choice of parameters as those reported in Ref.\,\cite{Hashimoto:2014gta}.
Note that if we start from the overlap-Dirac operator, which is realized in the limit of $N_7 \to \infty$ and it has an exact sign function, the theory becomes nonlocal and phenomenologically unnatural.

To introduce magnetic flux background, we assume that link functions $Q_M(y_5, y_6)$ in the toroidal directions acquire the following expectation values:
\begin{align}
Q_5(y_5,y_6) &=
\begin{cases}
1 & \quad (y_5 \neq 1),\\
\exp[-i b y_6] & \quad (y_5= 1),\\
\end{cases}
\\
Q_6(y_5, y_6) &= \exp[i b y_5],\label{eq:def_mag_u} 
\end{align}
where $y_M \in \{1/N_M,2/N_M, \cdots,1 \} \,\, (M=5, 6)$, and $b = 2\pi m$ is required from the consistency of single particle wavefunction \cite{AlHashimi:2008hr}.

\vspace{5pt}
{\em (De)Constructed zero-mode wavefunctions.}---
In order to obtain zero-mode wavefunctions, the main task is to solve an eigenvalue problem,
\begin{align}
H_\text{DW} \psi_n^{(D)} (y) = \lambda_n \psi_n^{(D)} (y),
\label{eq:eigenvalue_eq}
\end{align}
where $y \equiv (y_5,y_6)$. 
Eigenvalues and eigenvectors in the eigenvalue problem correspond to the KK-decomposed mass spectra and mode functions.
To realize Gaussian-localized zero-modes, it is necessary to select appropriate bases out of eigenvectors in Eq.\,\eqref{eq:eigenvalue_eq} such that all of kinetic terms are canonically normalized.
This is because, on the lattice, there is no reason to be eigenstates of a covariant translational operator $\tilde{Y}$ defined in Ref.\,\cite{Abe:2014noa}. 
Note that the zero-mode wavefunctions in Ref.\,\cite{Cremades:2004wa} are all eigenstates of the operator $\tilde Y$.

\begin{figure}[tb]
\centering
\includegraphics*[width=0.375\textwidth]{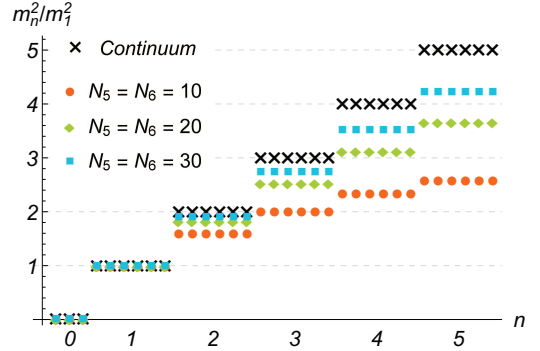}
\caption{The KK spectra in the continuum (black cross symbols) and (de)constructed (colored dots) theories, where the square (rhombus and circle) symbols denote those of $N_5=N_6=30$ ($20$ and $10$), respectively.}
\label{fig:kkmass}
\end{figure}

\begin{figure*}[tb]
\centering
\includegraphics[width=0.3\textwidth]{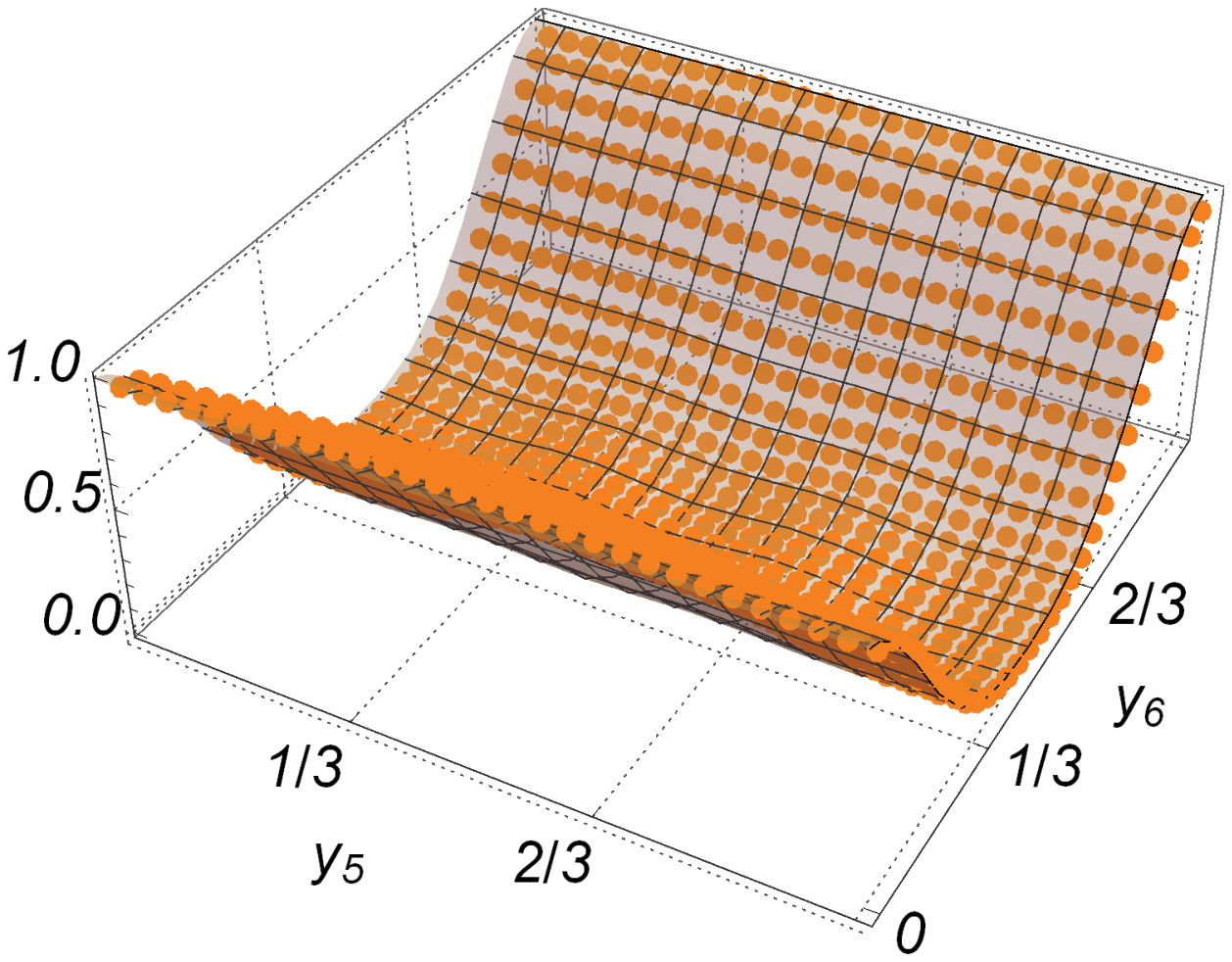}\hspace{0.033\textwidth}
\includegraphics[width=0.3\textwidth]{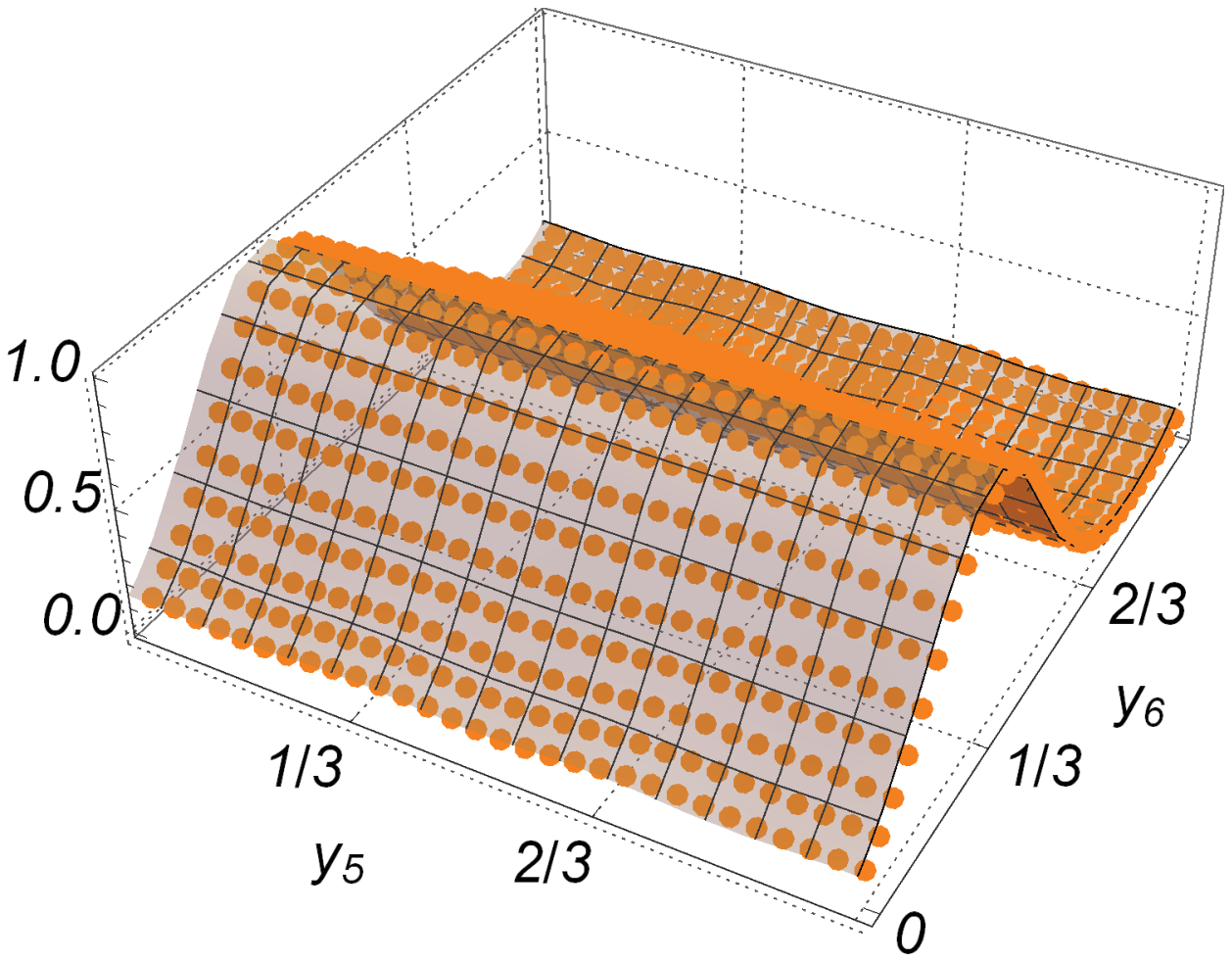}\hspace{0.033\textwidth}
\includegraphics[width=0.3\textwidth]{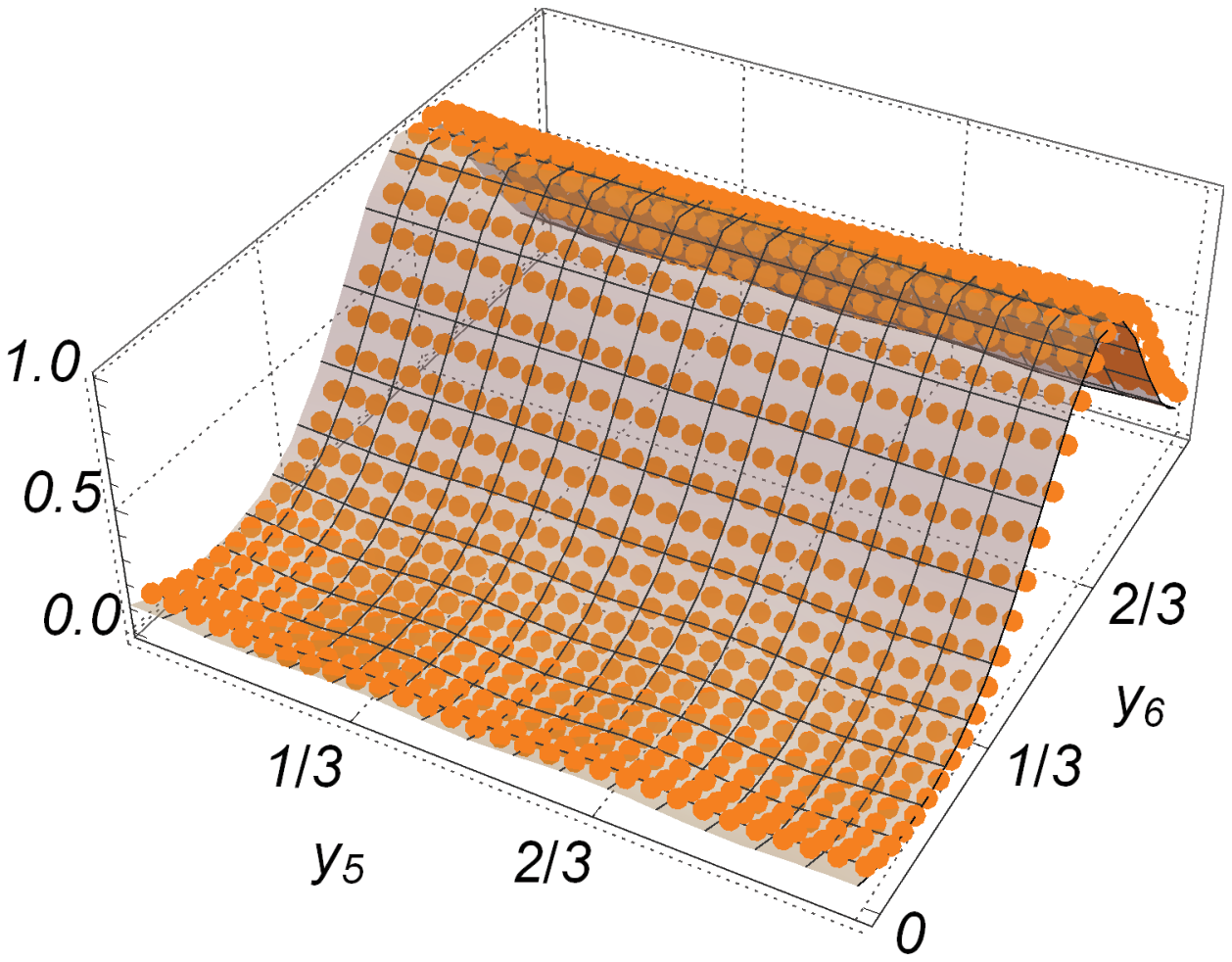}
\caption{Zero-mode wavefunctions for Eq.\,\eqref{eq:eigenvalue_eq} (orange dots) and those in the continuum theory \eqref{eq:eigenvalue_CIM_eq} (shaded curved surfaces), where we set $N_5=N_6=30$.}
\label{fig:zero-mode-wavefunctions-ma3}
\end{figure*}

In our practical calculation, we take the number of lattice points in the seventh direction is sixteen, i.e., $N_7 = 16$, and set $N_5=N_6=30$, $20$ and $10$ for comparison.
Also, we fix the number of magnetic fluxes as $m=3$.
Our code is implemented by Python 3.4 and Numpy with Cython from the scratch, and the calculation is performed in our laptop and desktop computers.
By the setup previously shown, it is possible to realize localization profiles of the KK wavefunctions as well as several lower modes of the KK mass spectra $m_n^2=4 \pi m n$ \eqref{eq:KK_mass} as shown in Fig.\,\ref{fig:kkmass}.
It is remarkable that negative chirality modes never appear in the lowest eigenvalues, as recognized in the continuum theory, and thus the (de)construction mechanism also lead to chirality projection via the presence of nonvanishing fluxes (monopoles).
Note that the degeneracy of each of KK levels is controlled by $N_7$, while the mass squared ratios of the KK spectra are determined by $N_5$ and $N_6$.
The deviation between the continuum and (de)constructed theories can be recognized, depending on the numbers of $N_5$ and $N_6$.
Next, we depict the (de)constructed zero-mode wavefunctions $\psi^{(D), i}_0 \,\, (i=0, 1, 2)$ in Fig.\,\ref{fig:zero-mode-wavefunctions-ma3}, where one can easily find satisfactory agreements between each of zero-mode wavefunctions for Eq.\,\eqref{eq:eigenvalue_eq} and those in the continuum theory \eqref{eq:eigenvalue_CIM_eq}.
It is also found that the scalar components are discretely realized in a similar manner.
Thus, we conclude that the magnetized extra dimensions can be (de)constructed.

\vspace{5pt}
{\em Conclusion and discussion.}---
In this Letter, we have reconstructed the toroidal compactification with magnetic fluxes on the basis of an idea by Arkani-Hamed et al., {\em (de)construction}.
As a main result, we have concretely established the renormalizable, asymptotically free, four dimensional local gauge theories where family replications, e.g., three-generation structures of the SM fermions, originate from nonvanishing magnetic fluxes in the toroidal compactification.
Because the toroidal compactification with fluxes possesses nontrivial topological index, it turned out that the (de)construction of such a situation is required to utilize the three dimensional Moose diagram with the Wilson fermion, namely the domain-wall fermion.
In this case, although the total topological index is zero, contributions from the zero-moodes and the heaviest KK modes (i.e., doubler modes) cancel out each other.
As long as we focus on the lower modes, nontrival topological index is effectively realized for the 6D theory.
In addition, the localization profiles of multiple zero-mode wavefunctions in the continuum theory have been realized by the multiplicity of the latticized gauge theories.

Although it is not mentioned in the main text, we can reproduce effective coupling constants, especially (three-point) Yukawa coupling constants, after dimensional reduction.
If we focus on the lowest modes among the KK-decomposed modes, the 6D fields are expressed as 
$\Psi(x^\mu, z) = \sum_j \chi^j_{0}(x^\mu) \psi^j_{+, 0}(z)$ and $\Phi(x^\mu, z) = \sum_j \varphi^j_{0}(x^\mu) \phi^j_{+, 0}(z)$, respectively.
Then, integrating out 6D Yukawa interaction terms, e.g., the interaction among quarks and the Higgs boson in the SM, along two extra dimensions is expressed as
\begin{align}
{\cal L}_{\rm Yukawa} &\propto \int_{T^2} d^2z \, \Psi(x^\mu, z) \bigl( \Psi(x^\mu, z) \bigr)^\dag \Phi(x^\mu, z) \notag\\
&= \left(\int_{T^2} d^2z \, \psi^i_{+, 0}(z) \bigl( \psi^j_{+,0}(z) \bigr)^\dag \phi^k_{0}(z) \right) \notag\\
& \hspace{60pt} \times \chi^i_{0} (x^\mu) \bigl( \chi^j_{0}(x^\mu) \bigr)^\dag \varphi_{0}^k(x^\mu).
\end{align}
Thus, the effective Yukawa couplings are calculated by overlap integrations of three kinds of the zero-mode wavefunctions \cite{Cremades:2004wa}:
\begin{gather}
y_{ijk} = \int_{T^2} d^2z \, \psi_{+, 0}^i(z) \bigl( \psi_{+, 0}^j(z) \bigr)^\dag \phi_0^k(z),
\end{gather}
up to an overall factor.
Since the zero-mode wavefunctions succeeded to be (de)constructed, the corresponding Yukawa coupling constants should be calculated as follows:
\begin{gather}
y^{(D)}_{ijk} = \sum_{y} \psi^{(D), i}(y) \bigl( \psi^{(D), j} (y) \bigr)^* \phi^{(D), k}(y).
\end{gather}
Detailed analyses of the (de)constructed Yukawa couplings are left for our future work.
In particular, corrections from small values of $N_5, N_6$ and $N_7$ to Yukawa couplings are characteristic features of our setup.
Those might be detected in the future experiments \cite{Dawson:2013bba}.

As a related topic, the lattice quantum chromodynamics with magnetic fluxes is actively investigated in the context of the relativistic heavy ion collision at LHC and RHIC.
Elliptic flows of the quark gluon plasma generate the hugest magnetic field in the world, and thus researches of the equation of states and phase transition with external magnetic fluxes have attracted much attentions \cite{DElia:2010abb, Bali:2011qj, Endrodi:2015oba}.
A crucial difference between such researches and ours appears in a coupling to gluons, which makes the system under consideration more intricate, and for example they derive chiral symmetry breaking.
For this reason, our setup is not related to such researches directly.
However, our research in this Letter provides a great insight of the role of chiral symmetry on the lattice in the presence of the external magnetic fluxes.

It should be mentioned to a relation between our model and an effective Hamiltonian of 3D topological insulators (TI) \cite{Zhang:2009zzf, Araki:2013qva}.
As introduced previously, our Lagrangian is the Wilson fermion in the 3D, and is coincident with the effective Hamiltonian of 3D TI.
Recently, surface states of 3D TI with the external magnetic field have been observed in an experiment \cite{Xu:2014}.
It would indicate that the physics of 3D TI is the same as the extra dimensional theory which can predict family replications of elementary particles.

\vspace{5pt}
{\em Acknowledgment.}---
The authors would like to thank Hiroyuki Abe, Yutaka Hosotani, Tetsuya Onogi, Hooman Davoudiasl, Taku Izubuchi, Hidenori Fukaya and Issaku Kanamori for helpful comments.
Y.T. would like to sincerely thank Masafumi Kurachi for valuable discussions at each stage of this research.
Y.T. is supported in part by Grants-in-Aid for Scientific Research No.\,16J04612 from the Ministry of Education, Culture, Sports, Science and Technology in Japan.
A.T. is supported by NSFC under Grant No.\,11535012.


\begin{thebibliography}{99}
\bibitem{Bachas:1995ik}
  C.~Bachas,
  hep-th/9503030.


\bibitem{Cremades:2004wa}
  D.~Cremades, L.~E.~Ibanez and F.~Marchesano,
  JHEP {\bf 0405} (2004) 079
  [hep-th/0404229].


\bibitem{ArkaniHamed:2001ca}
  N.~Arkani-Hamed, A.~G.~Cohen and H.~Georgi,
  Phys.\ Rev.\ Lett.\  {\bf 86} (2001) 4757
  [hep-th/0104005].


\bibitem{Atiyah:1963zz} 
  M.~F.~Atiyah and I.~M.~Singer,
  Bull.\ Am.\ Math.\ Soc.\  {\bf 69}, 422 (1969).


\bibitem{Nielsen:1981hk} 
  H.~B.~Nielsen and M.~Ninomiya,
  Phys.\ Lett.\  {\bf 105B}, 219 (1981).


\bibitem{Hamada:2012wj} 
  Y.~Hamada and T.~Kobayashi,
  Prog.\ Theor.\ Phys.\  {\bf 128}, 903 (2012)
  [arXiv:1207.6867 [hep-th]].


\bibitem{Tomiya:2016jwr} 
  A.~Tomiya, G.~Cossu, S.~Aoki, H.~Fukaya, S.~Hashimoto, T.~Kaneko and J.~Noaki,
  arXiv:1612.01908 [hep-lat].


\bibitem{Brower:2005qw} 
  R.~C.~Brower, H.~Neff and K.~Orginos,
  Nucl.\ Phys.\ Proc.\ Suppl.\  {\bf 153}, 191 (2006)
  [hep-lat/0511031].


\bibitem{Fukaya:2004kp} 
  H.~Fukaya and T.~Onogi,
  Phys.\ Rev.\ D {\bf 70}, 054508 (2004)
  [hep-lat/0403024].


\bibitem{Brower:2012vk} 
  R.~C.~Brower, H.~Neff and K.~Orginos,
  arXiv:1206.5214 [hep-lat].


\bibitem{Kaplan:2009yg} 
  D.~B.~Kaplan,
  arXiv:0912.2560 [hep-lat].


\bibitem{Hashimoto:2014gta} 
  S.~Hashimoto, S.~Aoki, G.~Cossu, H.~Fukaya, T.~Kaneko, J.~Noaki and P.~A.~Boyle,
  PoS LATTICE {\bf 2013}, 431 (2014).


\bibitem{AlHashimi:2008hr} 
  M.~H.~Al-Hashimi and U.-J.~Wiese,
  Annals Phys.\  {\bf 324}, 343 (2009)
  [arXiv:0807.0630 [quant-ph]].


\bibitem{Abe:2014noa} 
  T.~h.~Abe, Y.~Fujimoto, T.~Kobayashi, T.~Miura, K.~Nishiwaki and M.~Sakamoto,
  Nucl.\ Phys.\ B {\bf 890}, 442 (2014)
  [arXiv:1409.5421 [hep-th]].


\bibitem{Dawson:2013bba} 
  S.~Dawson {\it et al.},
  arXiv:1310.8361 [hep-ex].


\bibitem{DElia:2010abb} 
  M.~D'Elia, S.~Mukherjee and F.~Sanfilippo,
  Phys.\ Rev.\ D {\bf 82}, 051501 (2010)
  [arXiv:1005.5365 [hep-lat]].


\bibitem{Bali:2011qj} 
  G.~S.~Bali, F.~Bruckmann, G.~Endrodi, Z.~Fodor, S.~D.~Katz, S.~Krieg, A.~Schafer and K.~K.~Szabo,
  JHEP {\bf 1202}, 044 (2012)
  [arXiv:1111.4956 [hep-lat]].


\bibitem{Endrodi:2015oba} 
  G.~Endrodi,
  JHEP {\bf 1507}, 173 (2015)
  [arXiv:1504.08280 [hep-lat]].


\bibitem{Zhang:2009zzf} 
  H.~Zhang, C.~X.~Liu, X.~L.~Qi, X.~Dai, Z.~Fang and S.~C.~Zhang,
  Nature Phys.\  {\bf 5}, 438 (2009).


\bibitem{Araki:2013qva} 
  Y.~Araki, T.~Kimura, A.~Sekine, K.~Nomura and T.~Z.~Nakano,
  PoS LATTICE {\bf 2013}, 050 (2014)
  [arXiv:1311.3973 [cond-mat.str-el]].


\bibitem{Xu:2014}
  Y.~Xu, I.~Miotkowski, C.~Liu, J.~Tian, H.~Nam, N.~Alidoust, J.~Hu, C.~K.~Shih, M.~Z.~Hasan and Y.~P.~Chen
  Nature Phys.\  {\bf 10}, 956 (2014).
\end{thebibliography}

\end{document}